\documentclass[useAMS,usenatbib]{mn2e}
\usepackage{graphicx,amssymb,color}
\usepackage[normalem]{ulem}
\newcommand{\rev}{ }

\title[Evolved planetary magnetospheres]
{Planetary magnetosphere evolution around post-main-sequence stars}

\author[]{Dimitri Veras$^{1,2}$\thanks{E-mail: d.veras@warwick.ac.uk}\thanks{STFC Ernest Rutherford Fellow},
Aline A. Vidotto$^{3}$
\\
$^{1}$Centre for Exoplanets and Habitability, University of Warwick, Coventry CV4 7AL, UK
\\
$^{2}$Department of Physics, University of Warwick, Coventry CV4 7AL, UK
\\
$^{3}$School of Physics, Trinity College Dublin, the University of Dublin, Dublin-2, Ireland
}

\pubyear{2021}

\begin{document}
\label{firstpage}
\pagerange{\pageref{firstpage}--\pageref{lastpage}}
\maketitle

\begin{abstract}
Accompanying the mounting detections of planets orbiting white dwarfs and giant stars are questions about their physical history and evolution, particularly regarding detectability of their atmospheres and potential for habitability. Here we determine how the size of planetary magnetospheres evolve over time from the end of the main sequence through to the white dwarf phase due to the violent winds of red giant and asymptotic giant branch stars. By using a semianalytic prescription, we investigate the entire relevant phase space of planet type, planet orbit and stellar host mass ($1-7M_{\odot}$). We find that a planetary magnetosphere will always be quashed at some point during the giant branch phases unless the planet's magnetic field strength is at least two orders of magnitude higher than Jupiter's current value. We also show that the time variation of the stellar wind and density generates a net increase in wind ram pressure and does not allow a magnetosphere to be maintained at any time for field strengths less than $10^{-5}$ T (0.1 G). This lack of protection hints that currently potentially habitable planets orbiting white dwarfs would have been previously inhospitable.  
\end{abstract}


\section{Introduction}

To date, more than 4,000 exoplanets have been identified. Although the majority of the detected exoplanets orbit main-sequence stars, we now know of over 100 planets which orbit red giant stars \citep[e.g.,][]{refetal2015,gruetal2019,witetal2020}\footnote{www.lsw.uni-heidelberg.de/users/sreffert/giantplanets} and four planets which orbit white dwarfs \citep{thoetal1993,sigetal2003,luhetal2011,ganetal2019,vanetal2020}.

All these planets formed during the protoplanetary disc phase \citep[e.g.,][]{pinetal2018} and have since survived stellar evolution to reach the present time. These evolved planets provide valuable benchmarks when tracing the full lifetime of their parent systems \citep{veras2016}, and, when combined with additional data, can help constrain formation locations in the disc \citep{haretal2018,veretal2020} and link system chemistry with architecture \citep{payetal2016,payetal2017,xuetal2017,musetal2018,swaetal2019,doyetal2021}.

After the main-sequence phase, stars in the mass range $1-7 M_\odot$ become red giant branch (RGB) stars, ascend through the asymptotic giant branch (AGB), and then traverse the planetary nebulae phase, eventually ending their lives as white dwarfs. Accompanying the non-monotonic and often sudden changes in radii and surface temperatures (or luminosities) are dramatic variations in stellar wind properties. 

For example, Solar-mass stars have winds that are rarefied and decreasing throughout main-sequence evolution \citep{vidotto2021}; the Sun's current mass loss rate is $2\times 10^{-14}M_\odot/$yr. However, these winds become denser and more violent as they evolve off the main sequence, with peak values of $\approx 10^{-7}-10^{-5}M_\odot/$yr \citep{reimers1977,vaswoo1993,catelan2000,wooetal2016,ofietal2021}.

\begin{figure*}
\centerline{
\includegraphics[width=17cm]{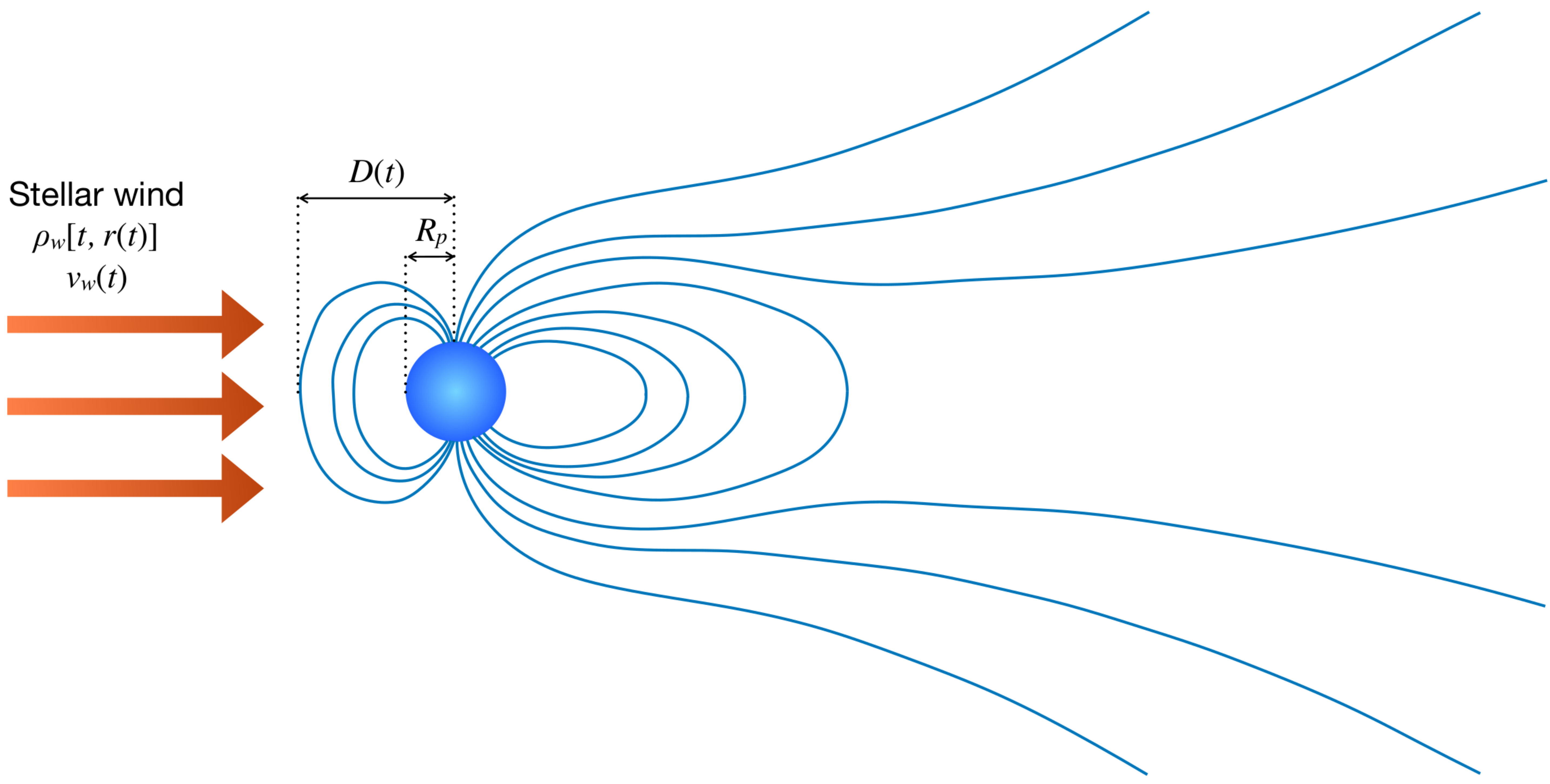}
}
\caption{
Schematic illustrating the geometric meaning of the magnetopause distance $D(t)$ and the radius of the planet $R_{\rm p}$. Both the velocity $v_{\rm w}(t)$ and density $\rho_{\rm w}[t,r(t)]$ of the stellar wind are time-dependent.
}
\label{Defs2}
\end{figure*}

As a result, and because stellar winds permeate the interplanetary medium, this region becomes more dense during post-main-sequence evolution. The external environments around all orbiting planets become altered. A further effect of post-main-sequence evolution is that the terminal velocity of the stellar wind shrinks. For comparison, the present-day solar wind reaches terminal velocities of around $400-800$ km/s \citep{mccetal2008}. Although the terminal velocities of RGB and AGB stars can be significantly lower ($\lesssim 100$ km/s, \citealt{wooetal2016}), the substantial increase in wind density still leads to an overall higher wind ram pressure beyond the main sequence (as we will demonstrate in this paper).

Further, the stronger winds of RGB and AGB stars can harm prospects for planetary habitability. High stellar wind ram pressure can more easily erode planetary atmospheres, which are believed to be key for planetary habitability. Indeed, studies featuring post-main-sequence habitability prominently feature planetary atmospheres \citep{agol2011,fosetal2012,barhel2013,ramkal2016,kozetal2018,kozkal2019,kozkal2020,kozetal2020}. The direct interaction between planetary atmospheres and stellar winds, which can lead to atmospheric erosion \citep[see, e.g., the case of young Mars in][]{kuletal2007}, can be avoided if the planet harbours a large-scale (intrinsic) magnetic field. 

Intrinsic magnetospheres are therefore believed to be key factors in establishing the habitability of a planet \citep[e.g.,][]{tacetal2011,videtal2013}. The mounting interest in characterising habitability in extrasolar systems has motivated dedicated studies on exoplanetary magnetospheres at an increasing pace \citep[e.g.,][]{gunetal2018,viletal2018,caretal2019,zhibis2019,turetal2020,basnan2021,greetal2021}. However, such investigations have largely focussed on main-sequence or pre-main-sequence $\lesssim 1.2M_{\odot}$ stars. 

In this work, we focus on the evolution of planetary magnetospheres {\it after} their host stars have evolved off the main sequence. We consider planet-host stars with masses $1-7 M_\odot$. Our stellar evolution model accounts for the presence of stellar winds, allowing us to calculate the evolution of stellar wind ram pressures at any orbital distance. We place fictitious planets orbiting their hosts at a range of initial distances, taking also into account planetary orbital expansion. The planetary magnetic field provides the necessary magnetic pressure, which, if sufficiently large, can prevent the wind from directly impacting the stellar surface \citep{chafer1930}. The stand-off distance $D(t)$ where the wind-magnetosphere interaction takes place is shown schematically in Fig. \ref{Defs2}. Our present study, therefore, allows us to determine whether, at any point in its evolution,  planetary magnetospheres become significantly small, thus possibly affecting atmospheric retention. Our work has important consequences for future studies on planetary habitability of evolved planetary systems.

\begin{figure*}
\centerline{
\includegraphics[width=8.5cm]{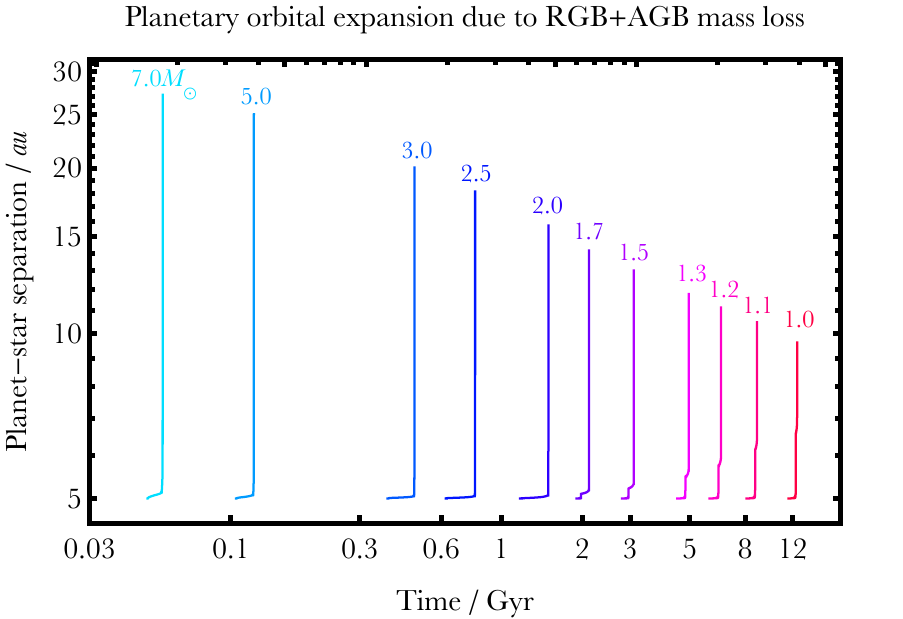}
\includegraphics[width=8.5cm]{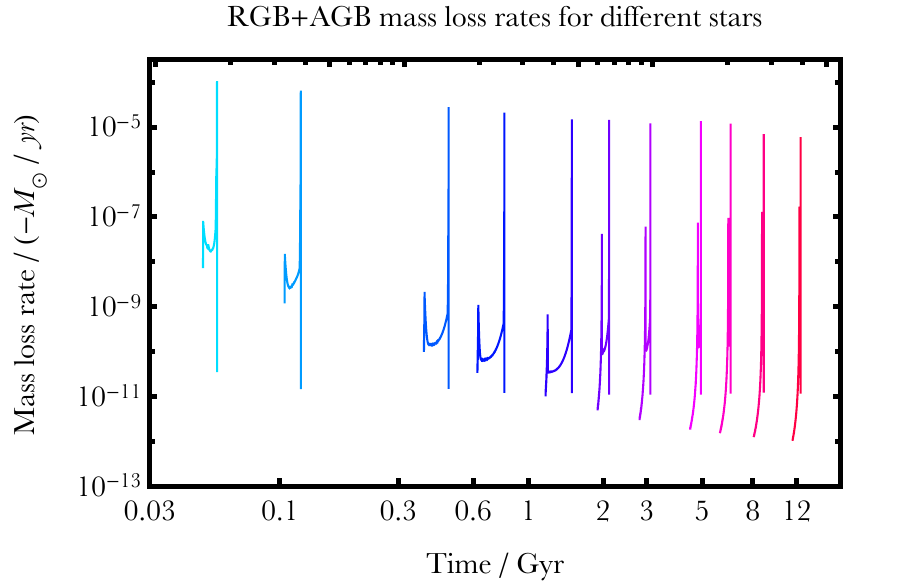}
}
\centerline{}
\centerline{
\includegraphics[width=8.5cm]{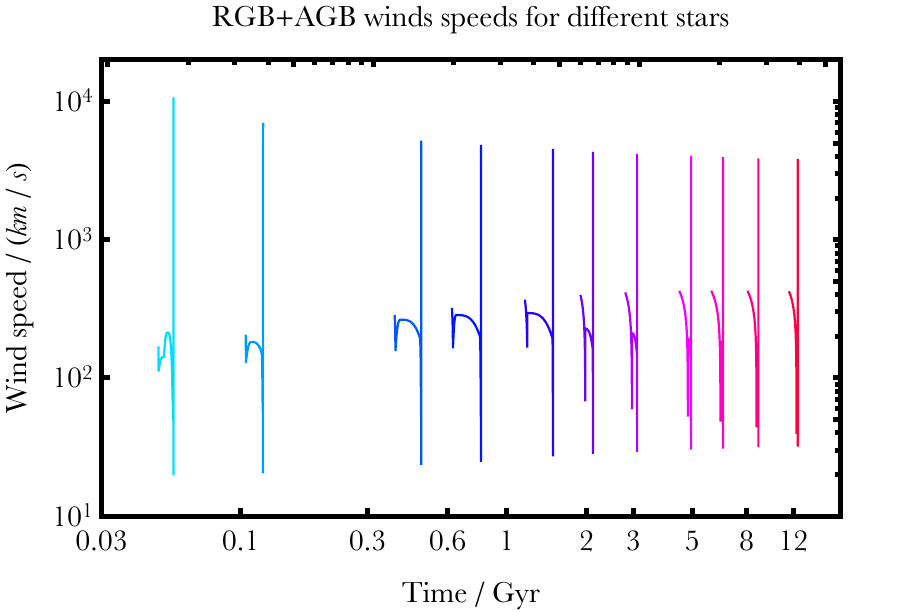}
\includegraphics[width=8.5cm]{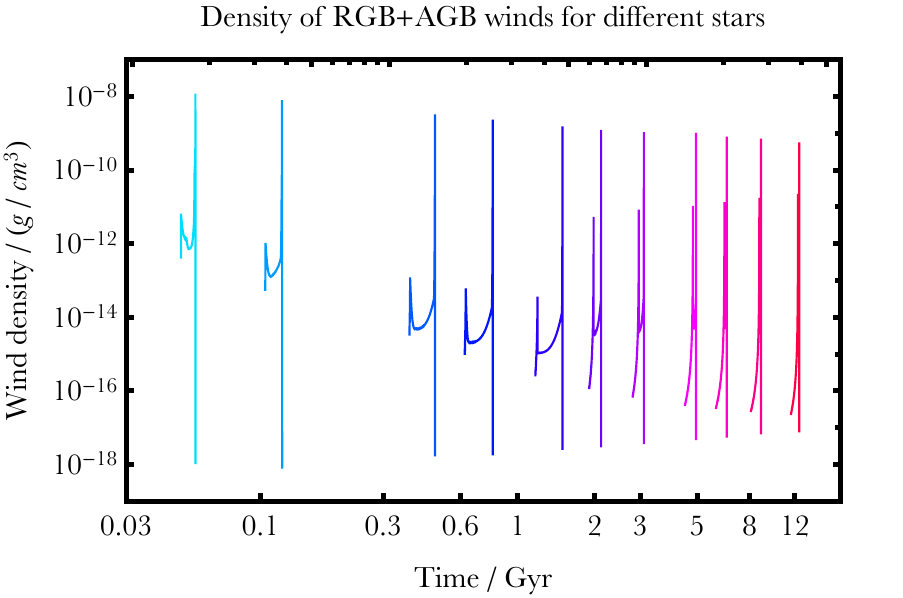}
}
\caption{
The evolution of different post-main-sequence planetary systems throughout the RGB and AGB phases only, including the planetary nebula transition to white dwarfs. Time is measured from the zero-age-main-sequence, such that the RGB phase occurs later for lower mass stars. The curves are partitioned according to initial stellar mass, which is given by the coloured numbers in the upper left panel. In that panel, the planet's separation at the beginning of the RGB phase is 5 au. The other panels illustrate the wind mass loss rate, wind speed and wind density. 
}
\label{Stellar}
\end{figure*}

\section{Computations}

Consider an arbitrary planet (e.g., terrestrial or giant) of radius $R_{\rm p}$ with a dipolar magnetic field with an equatorial field strength of $B_{\rm p}$, and with an atmosphere where the ram and thermal pressures are negligible. Both $R_{\rm p}$ and $B_{\rm p}$ are assumed to remain constant in time. 

\subsection{Basic equations}

Our goal is to estimate the size of its magnetopause, given by the standoff distance $D(t)$, whose geometric meaning is given in Fig. \ref{Defs2}. We can do so by balancing the ram pressure of the stellar wind and the magnetic pressure of the planet such that \citep{chafer1933,blatar2018,videtal2019}

\[
D(t)^3 = 2^{\frac{1}{3}} R_{\rm p}^3 \sqrt{
\frac
{B_{\rm p}^2}
{8 \pi \rho_{\rm w}\left[t,r(t)-D(t)\right] v_{\rm w}\left[t,r(t)-D(t)\right]^2}
}
\]

\begin{equation}
\ \ \ \ \ \ \ \  \approx 2^{\frac{1}{3}}\frac{R_{\rm p}^3 B_{\rm p}}{v_{\rm w}(t) }\sqrt{
\frac
{1}
{8\pi \rho_{\rm w}\left[t,r(t)\right]}
},
\label{standoff}
\end{equation}

\noindent{}where $r(t)$ is the planet-star separation, $v_{\rm w}$ is the speed of the wind and $\rho_{\rm w}$ is the density of the wind. In our approximation, we have assumed that $D(t) \ll r(t)$ and that the wind speed is independent of distance for the planet-star separations that we will sample. This assumption is reasonable for distances beyond tens of stellar radii, where the wind velocity becomes the terminal velocity. Henceforth, we consider $v_{\rm w}(t)$ to be the terminal velocity at time $t$.

We can estimate and express both $v_{\rm w}$ and $\rho_{\rm w}$ in terms of stellar parameters. Denote the host star's mass and radius as $M_{\star}(t)$ and $R_{\star}(t)$, and assume that the stellar radius is defined to always extend out to the edge of the envelope. The terminal velocity is on the order of the surface escape speed, such that

\begin{equation}
v_{\rm w}(t) = \alpha \sqrt{\frac{2 G M_{\star}(t)}{R_{\star}(t)}},
\label{wind}
\end{equation}

\noindent{}where $\alpha \approx 0.5-4.0$.

Provided that a stellar wind exists, the wind density can be estimated from the common assumption of spherical symmetry with

\begin{equation}
\rho_{\rm w}\left[t,r(t)\right] 
= -\frac{dM_{\star}(t)}{dt} 
\left(
\frac{1}{4 \pi r(t)^2 v_{\rm w}(t)}
\right).
\end{equation}

In this investigation, we consider only post-main-sequence phases of evolution. Giant branch stars have time-variable and sometimes strong winds, whereas white dwarfs have no winds. We assume that the stellar mass loss from the giant branch winds is isotropic, an approximation which is good enough for our purposes \citep{veretal2013}. {\rev This isotropy assumption allows for the system to conserve angular momentum by maintaining rotational symmetry through Noether's Theorem. In this way, although the system does not conserve energy due to mass loss, it does conserve angular momentum.}

{\rev Under this isotropic assumption, the response of a secondary body due to mass loss from the primary has been studied for over a century \citep{gylden1884,mestschersky1893}; the resulting complete equations of motion in terms of orbital elements were subsequently given by \cite{omarov1962} and \cite{hadjidemetriou1963}. However,} exoplanets reside close enough to their parent stars to allow us to use {\rev a simplified form of these equations due to} the adiabatic mass loss approximation for orbital expansion. 

{\rev The simplified form is obtained by applying conservation of angular momentum in conjunction with the assumption that the orbital eccentricity does not change due to mass loss. \cite{veretal2011} demonstrated the robustness of this approximation for planetary orbits within hundreds of au of their parent stars. The final result is} 

\begin{equation}
\frac{dr(t)}{dt} \approx -\frac{r(t)}{M_{\star}(t)} \frac{dM_{\star}(t)}{dt}
\end{equation}

\noindent{}which admits the solution

\begin{equation}
\frac{r(t)}{r(0)} \approx \frac{M_{\star}(0)}{M_{\star}(t)},
\label{drdt}
\end{equation}

\noindent{}where $t=0$ corresponds to the beginning of the giant branch phases of stellar evolution.

Having established these system properties, we now require stellar evolution profiles. We compute these profiles from the {\tt sse} code \citep{huretal2000} by assuming in all cases Solar metallicity, a Reimers mass loss coefficient of 0.5, and a superwind prescription according to \cite{vaswoo1993}. We sample 11 initial stellar masses from $1.0M_{\odot}$ to $7.0M_{\odot}$. We justify this range because sub-solar masses take over a Hubble time to evolve, and on the higher end investigating planetary systems with host star masses close to the supernova limit is becoming increasingly relevant \citep{veretal2020,holetal2021}.

In Fig. \ref{Stellar} we illustrate the evolution of the stellar wind mass loss rate, speed and density, as well as the resulting effect on a planet for all of the stellar evolution profiles. The plots in the figure illustrate that as the initial stellar mass increases, the RGB phase becomes less prominent and the AGB phase dominates. The spikes in the profiles are attributed to either the tip of the AGB, when the mass loss is greatest, or just after, during the planetary nebula transition, when the mass is largely fixed but the radius rapidly contracts into a white dwarf.

\begin{figure}
\centerline{
\includegraphics[width=8.5cm]{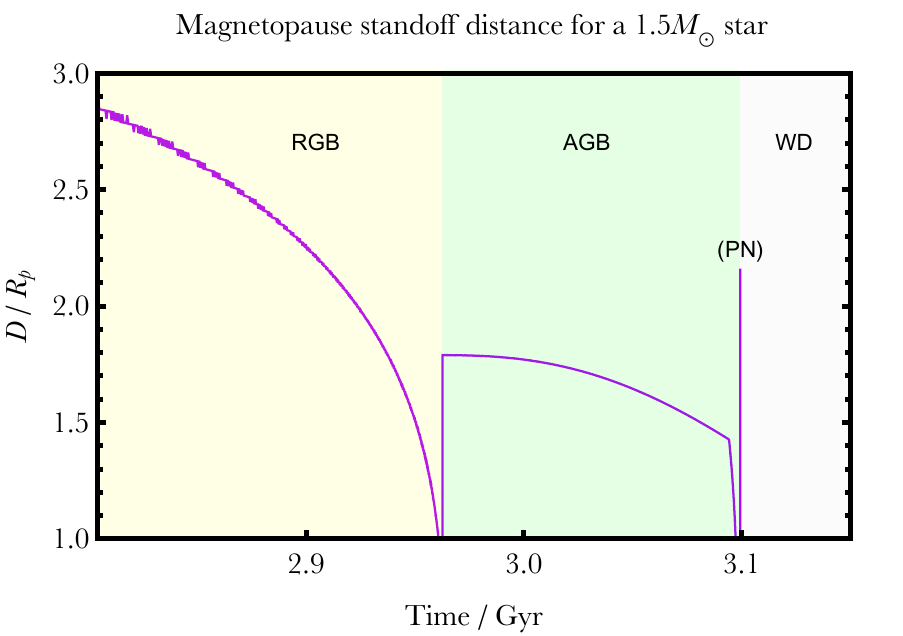}
}
\caption{
Example evolution profile of the planetary magnetopause standoff distance relative to the planet radius for a $1.5M_{\odot}$ star. Here we assume $B_{\rm p} = 10^{-3}$\,T, $r(0)=50$\,au and $\alpha=1$. During any part of the curve which lies below the $x$-axis (at the RGB and AGB tips), the magnetosphere disappears. These intervals each last for about 2 Myr. Soon after the planetary nebula (PN) spike, when the star becomes a white dwarf (WD), the stellar wind ceases. 
}
\label{DRplotind}
\end{figure}

\begin{figure}
\centerline{
\includegraphics[width=8.5cm]{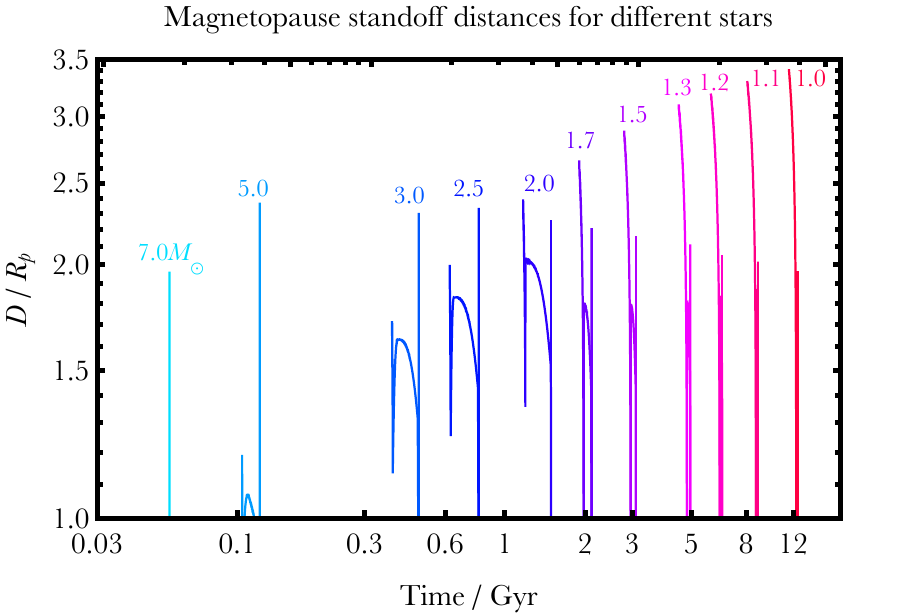}
}
\caption{
The planetary magnetopause evolution relative to planetary radius for all 11 stellar tracks, assuming $B_{\rm p} = 10^{-3}$\,T, $r(0)=50$\,au and $\alpha=1$. Like in Fig. \ref{Stellar}, here again time is measured from the zero-age-main-sequence. The highest spikes occur early on the RGB for the low-mass stars, whereas the highest spikes occur during the PN transition for the high-mass stars.
}
\label{DRplotgen}
\end{figure}

\begin{figure*}
\centerline{
\includegraphics[width=17cm]{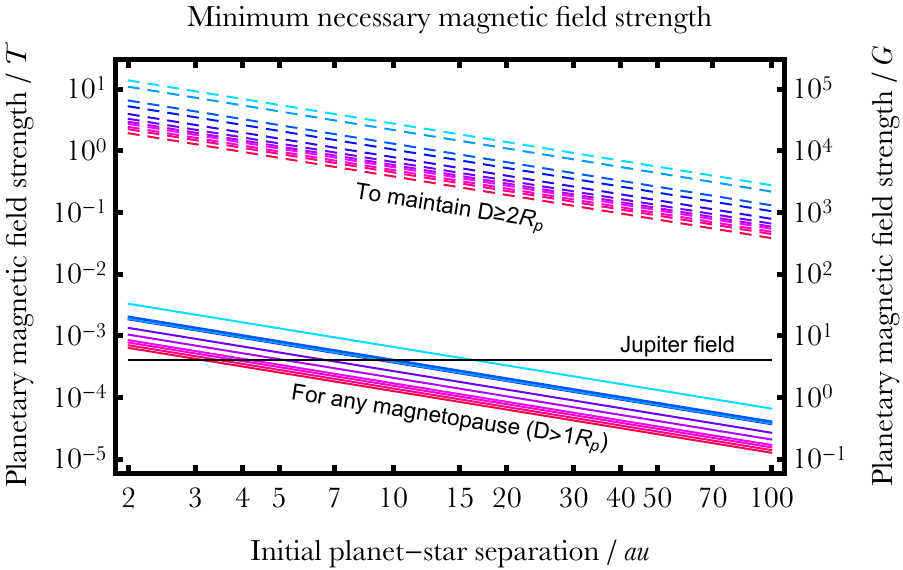}
}
\caption{
The minimum planetary magnetic field strengths necessary to satisfy certain conditions along the RGB and AGB phases. The first condition, given by the lower set of solid curves, is for a magnetopause to exist ($D > 1R_{\rm p}$) at any time. The second condition, given by the upper set of dashed curves, is for a field to maintain $D \ge 2R_{\rm p}$ throughout. In each set of curves, the bottom corresponds to $1.0M_{\odot}$ and the top corresponds to $7.0M_{\odot}$ (the colouring is consistent throughout the figures). The black horizontal line is Jupiter's current magnetic field strength of $\approx 4 \times 10^{-4}$\,T (4 G), the largest in the solar system. The high values of the magnetic fields required to provide protection suggest that planet atmospheres and surfaces will nearly always be exposed to erosive giant branch radiation. 
}
\label{Punch}
\end{figure*}

\subsection{Analysis}

Equations (\ref{wind})-(\ref{drdt}) allow us to solve for $D(t)$ (equation \ref{standoff}) with our minimal set of input parameters. Instead, however, we can reduce the parameter space to explore by computing the ratio $D(t)/R_{\rm p}$, and expressing that ratio with separated time-dependent stellar and fixed non-stellar components as

\begin{equation}
\frac{D(t)}{R_{\rm p}}
=
C f(t)
\label{geneq}
\end{equation}

\noindent{}where

\begin{equation}
f(t)
=
\frac
{R_{\star}(t)^{\frac{1}{12}}M_{\star}(0)^{\frac{1}{3}}}
{\dot{M}_{\star}(t)^{\frac{1}{6}} M_{\star}(t)^{\frac{5}{12}} }
.
\end{equation}

\noindent{}and

\begin{equation}
C
=
2^{-\frac{5}{36}}
\frac
{B_{\rm p}^{\frac{1}{3}} r(0)^{\frac{1}{3}}}
{\alpha^{\frac{1}{6}} G^{\frac{1}{12}}}
.
\label{cval}
\end{equation}

Here, the dot signals differentiation with respect to time, and again, $t=0$ corresponds to the beginning of the giant branch phases of stellar evolution. The value of $C$ is independent of time, and effectively scales the results for different stellar evolutions. 

As a demonstration of the evolution profile of the ratio $D/R_{\rm p}$, we plot this ratio for a $1.5M_{\odot}$ star in Fig. \ref{DRplotind}. In the plot, we assume $B_{\rm p} = 10^{-3}$\,T, $r(0)=50$\,au and $\alpha=1$. The values of $B_{\rm p}$ and $r(0)$ are high relative to solar system values, but are necessary to illustrate the evolution; lower values predominantly yield $D/R_{\rm p} < 1$, for which the magnetosphere is quenched. In this particular case, the magnetosphere is quenched for two intervals each lasting about 2 Myr, which could have implications for planetary protection.

Now we adopt these same values of $B_{\rm p}$, $r(0)$ and $\alpha$ for all stellar evolution profiles, and plot the result in Fig. \ref{DRplotgen}. Note that in this figure, the highest spikes for the lowest mass stars occur at the start of the RGB and for the highest mass stars at the planetary nebula transition, with the boundary occurring at around $2.0M_{\odot}$. Overall, the magnetopause distance varies significantly over the giant branch phases, regardless of stellar mass.

In order to usefully constrain the parameter space, we can now ask two questions: (i) can a planet host a magnetosphere at any point during the giant branch phases?, and (ii) how strong must the planetary magnetic field be in order to maintain, say $D/R_{\rm p} \ge 2$ throughout the giant branch phases? 

We can answer these questions through manipulation of $C$ and $f(t)$. The answer to the first question is given through a critical magnetic field strength by

\begin{equation}
B_{\rm p,crit1} = \frac{2^{\frac{5}{12}}G^{\frac{1}{4}}\alpha^{\frac{1}{2}}}{r(0) \left[ {\rm max}\left(f(t) \right) \right]^3}
\end{equation}

\noindent{}and the answer to the second question is given by

\begin{equation}
B_{\rm p,crit2} = \frac{2^{\frac{41}{12}}G^{\frac{1}{4}}\alpha^{\frac{1}{2}}}{r(0) \left[ {\rm min}\left(f(t) \right) \right]^3}.
\end{equation}

We plot these critical values in Fig. \ref{Punch} as a function of $r(0)$, assuming $\alpha = 1$. The value of $r(0)$ is bounded from below at about 2 au, which represents the approximate smallest planetary tidal engulfment distance along the giant branch phases \citep{musvil2012,adablo2013,norspi2013,viletal2014,madetal2016,prietal2016,ronetal2020}. We set an upper boundary at 100 au, the approximate conservative value for protoplanetary disc outer boundaries \citep{andrews2020}. 

The plot shows that an exo-Jovian analogue would just reach the threshold for hosting a magnetopause at some point during giant branch evolution. However, much higher fields would be required to maintain any magnetopause throughout these giant branch phases. For terrestrial and potentially habitable planets, any protection previously afforded by the magnetosphere would effectively disappear. This lack of protection, compounded with orbital expansion and varying stellar luminosities, suggest that life would be challenged to survive throughout the giant branch phases of stellar evolution.




\begin{figure}
\centerline{
\includegraphics[width=8.5cm]{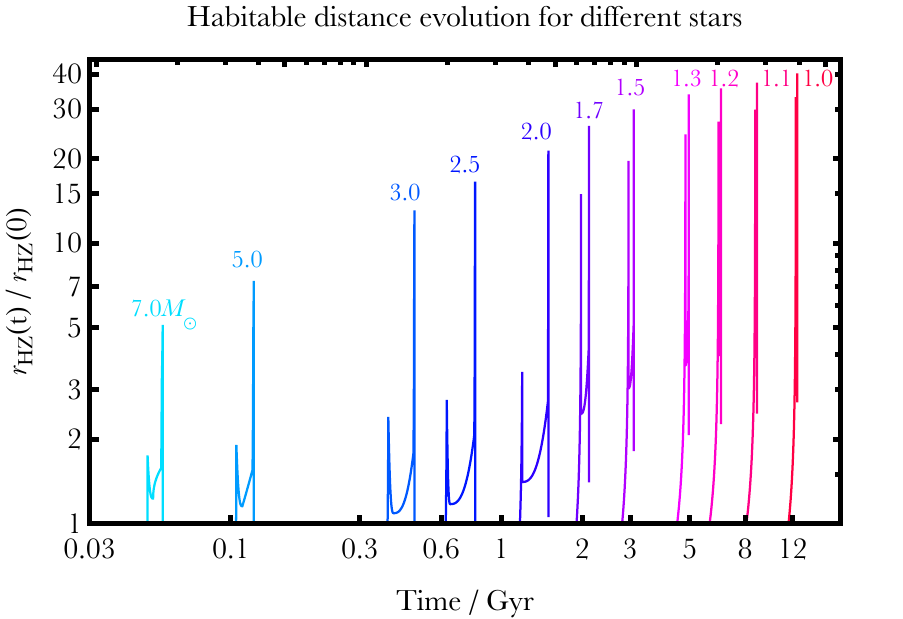}
}
\caption{
How the habitable distance expands and contracts due to giant branch evolution, from the simple model of Eq. (\ref{HZeq}). Increasing stellar mass reduces the level of expansion because of the high main-sequence baseline luminosities of the highest mass stars. In nearly all cases, the expansion of the habitable zone (here represented by a single distance) is faster than the expansion of the planetary orbit.
}
\label{HZplotgen}
\end{figure}

\section{Discussion}

Our treatment is simplistic and neglected additional effects which we know can be important, at least within the solar system. Jupiter's magnetosphere is inflated by plasma pressure, and there is a thermal pressure component in addition to the dynamic pressure component \citep{piletal2015}. However, unlike for the solar system planets, our knowledge of individual exoplanets is relatively small, and how e.g. plasma pressure inflates exo-magnetospheres is currently unknown.

{\rev Our formalism also assumed that throughout the giant branch phases, the planet's atmosphere did not undergo significant changes due to geological cycles or cometary impacts. Both of these potential triggers for atmospheric changes have received little attention in evolved planetary systems. If geological cycles re-form atmospheres slowly relative to post-main-sequence evolution durations, and cometary impacts in giant branch systems are infrequent enough \citep[Fig. 1 of][]{veretal2014} to render secondary atmosphere formation \citep{kraetal2018} ineffectual, then the maximum ram pressure experienced by the planet alone may act as a useful proxy for determining the fate of its magnetosphere.}

We have shown that, if the solar system planets are typical, then most exoplanets would not be able to retain a magnetosphere throughout the giant branch phases. The consequences for planetary protection are not immediately clear because, during these phases, the habitable zone shifts significantly. We can provide a very rough estimation of this shift through stellar evolution alone, and by considering a single distance rather than a zone. Assume that a planet is habitable at a distance of $r_{\rm HZ}(0)$ with a particular albedo and (habitable) equilibrium temperature. If this planet were to retain this same albedo and equilibrium temperature throughout its post-main-sequence evolution, then it would need to shift its position along the habitable distance according to, from Eq. (2) of \cite{valetal2014},

\begin{equation}
\frac{r_{\rm HZ}(t)}{r_{\rm HZ}(0)}
=
\sqrt{\frac{L_{\star}(t)}{L_{\star}(0)}}
\label{HZeq}
\end{equation}

\noindent{}where $L_{\star}$ is the stellar luminosity. This shift does not represent the actual movement of the planet, but rather the required shift to stay potentially habitable. 

By using $L_{\star}(t)$ values from the {\tt SSE} code \citep{huretal2000}, we plot the expansion of the habitable distance in Fig. \ref{HZplotgen}. The figure illustrates, perhaps counterintuitively, that the higher the stellar mass, the smaller the expansion of the habitable distance. This trend is due to the high main-sequence luminosities of the high mass stars: the $t=0$ luminosity baseline is already high. Further, in all cases, this habitable distance never exceeds a factor of 40 of its final main-sequence value. 

When compared with the actual orbital expansion of a planet (a factor of 2-5), the habitable zone always extends outward more quickly than a planet, except perhaps in the $7M_{\odot}$ case. This comparison, combined with the difficulty of retaining a magnetosphere throughout the giant branch phases of evolution, place doubt on whether a planet could remain habitable from the main sequence to the white dwarf phases.

Finally, we now consider the end products of stellar evolution. Because white dwarfs have no winds, around those stars ram pressure would not confine a planetary magnetosphere. If the white dwarf is sufficiently magnetic -- about 20 per cent of white dwarfs have detectable fields with $B_{\star} \gtrsim 0.1$~T (1~kG) and over 10 per cent have $B_{\star} \gtrsim 100$~T (1~MG) \citep{feretal2015,holetal2015,lanbag2019} -- then that field alone could provide the external pressure necessary to balance out the magnetic pressure of a planet. 

In contrast, neutron stars -- the end products of stars with masses of $7-20M_{\odot}$ -- do appear to emit winds \citep[e.g.][]{bluetal2017,enoetal2019,hsietal2021}. The three planets known to orbit single pulsars were in fact the first three confirmed exoplanets \citep{wolfra1992,wolszczan1994}, but are now known to be genuinely rare (as opposed to white dwarf planetary systems, which are common with an occurrence rate of 25-50 per cent; \citealt*{koeetal2014}). Further, those three pulsar planets are probably not in the habitable zone \citep{patkam2017} and are likely second-generation in the sense of having formed after the supernova \citep{milham2001}.

\section{Summary}

We have modelled the size evolution of planetary magnetospheres throughout the giant branch phases of evolution for 11 different stellar masses that encompass the entire range of white dwarf precursors. Our formalism is simple enough to be applied to both terrestrial and giant planets, and requires one to specify only the planet's magnetic field strength and initial separation (Eqs. \ref{geneq}-\ref{cval}). We have illustrated the magnetopause standoff distance evolution in Figs. \ref{DRplotind}-\ref{DRplotgen}, which suggest that most exoplanets will struggle to retain their magnetosphere (Fig. \ref{Punch}) until they reach the white dwarf phase. When combined with failing to keep up with the outward drift of the habitable zone (Fig. \ref{HZplotgen}), this lack of protection indicates that habitability around white dwarfs must be newly established around those stars.

\section*{Acknowledgements}

{\rev We thank the reviewer for their helpful comments, which have improved the manuscript.}
DV gratefully acknowledges the support of the STFC via an Ernest Rutherford Fellowship (grant ST/P003850/1). AAV has received funding from the European Research Council (ERC) under the European Union’s Horizon 2020 research and innovation programme (grant agreement no. 817540, ASTROFLOW).

\section*{Data Availability}

The simulation inputs and results discussed in this paper are available upon reasonable request to the corresponding author.

\label{lastpage}
\end{document}